\documentclass[11pt,twoside]{article}
\usepackage{./asp2014}
\usepackage{aasmacros}
\usepackage{color}

\aspSuppressVolSlug
\resetcounters

\begin{document}

\title{From Megaparsecs To Milliparsecs: Galaxy Evolution \& Supermassive Black Holes with NANOGrav and the ngVLA}
\author{Stephen~R.~Taylor$^{1,2}$ and Joseph Simon,$^{2,1}$
\affil{$^1$Theoretical AstroPhysics Including Relativity (TAPIR) Group, MC 350-17, California Institute of Technology, Pasadena 91125, CA, USA; \email{srtaylor@caltech.edu.edu}}
\affil{$^2$Jet Propulsion Laboratory, California Institute of Technology, 4800 Oak Grove Drive, Pasadena 91109, CA, USA; \email{joseph.j.simon@jpl.nasa.gov}}}

\paperauthor{Sample~Author1}{Author1Email@email.edu}{ORCID_Or_Blank}{Author1 Institution}{Author1 Department}{City}{State/Province}{Postal Code}{Country}
\paperauthor{Sample~Author2}{Author2Email@email.edu}{ORCID_Or_Blank}{Author2 Institution}{Author2 Department}{City}{State/Province}{Postal Code}{Country}

\section{Scientific Goals}

Within the current paradigm of hierarchical cosmological structure formation, galaxy growth occurs through a continuous process of gas and dark matter accretion, interspersed with major and minor mergers \citep[e.g][]{1984Natur.311..517B}. The fact that most massive galaxies are expected to harbor a supermassive black hole (SMBH) \citep{kormendy1995, Magorrian1998} has led to the understanding that central SMBHs share a common evolutionary history with their host galaxies. The prevalence of major mergers and the rate of gas accretion, both of which can contribute to the growth of SMBH and galaxy alike, are, as of yet, poorly constrained across cosmic time. One avenue for understanding the relative roles of these mechanisms is the study of post-merger galaxies, which are expected to harbor a ``dual'' SMBH system composed of the two black holes from each of the merging galaxies. Understanding the dynamical evolution of SMBH binaries in the rich physical environment of post-merger galactic environments is key to solving the puzzles of how SMBHs and galaxies co-evolve, how SMBHs grow over cosmic time, and how (or even if) the final stages of SMBH mergers take place in galaxy centers.

Precision timing of Galactic millisecond pulsars over the last decade has allowed the North American Nanohertz Observatory for Gravitational Waves (NANOGrav, and other collaborations around the globe) to use Arecibo, GBT, and VLA observations to forge a network of kiloparsec-spaced ``clocks'' that can respond to transiting nanohertz gravitational waves (GWs) through induced Doppler shifts of radio-pulse arrival rates. The dominant source of these nanohertz GWs is likely to be a population of SMBH binaries that are formed naturally in post-merger galaxies. The SMBHs from each galaxy will sink to the center of the common merger remnant through interactions with the galactic gas, stars and dark matter, and eventually become gravitationally bound as binaries \citep{bbr80}.  Through continued interaction with the environment, the binaries' orbits will tighten and GW emission will become an increasingly dominant factor in their continued evolution.  After sufficiently strong environmental interactions the binaries will eventually reach milliparsec orbital separation, at which point their GWs will be emitted in the Pulsar Timing Array (PTA) band at roughly nanohertz frequencies.

NANOGrav seeks to understand the evolution of galaxies and massive black-holes through cosmic time, as well as the dynamics of massive black-hole binaries formed through galaxy mergers. These goals intersect with the ngVLA science case in various ways; the most pertinent being:

\begin{figure}
\center
  \includegraphics[width=0.9\columnwidth]{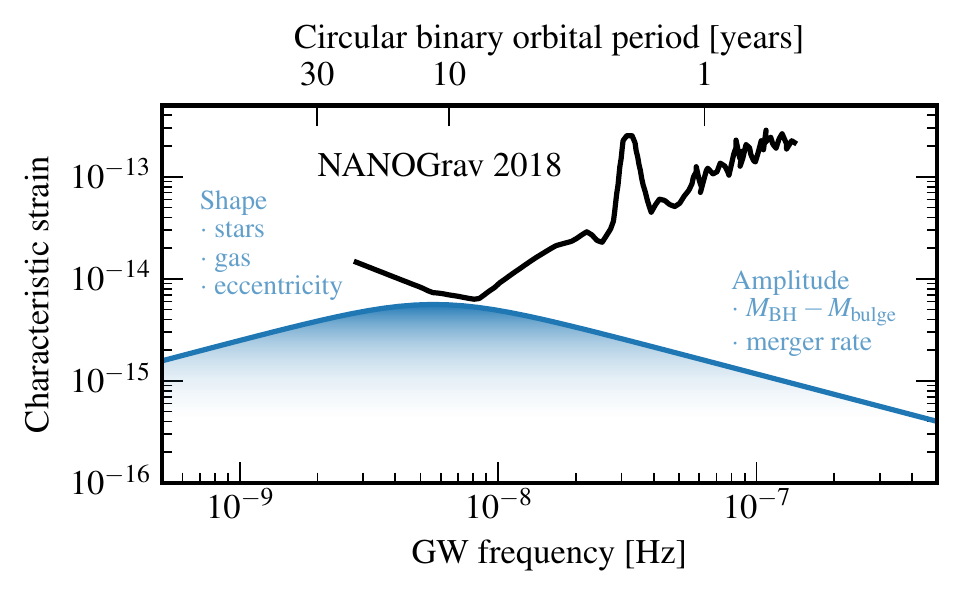}
  \caption{An illustration of the various astrophysical factors affecting the shape and amplitude of the characteristic strain spectrum of GWs emitted by a population of inspiraling supermassive black-hole binaries. This is contrasted against the per-frequency sensitivity of NANOGrav's most recent dataset, consisting of $45$ millisecond pulsars with precision timing over an $11$-year timescale. The spike at a frequency of $1\,\mathrm{yr}^{-1}$ corresponds to a loss in PTA sensitivity due to fitting each pulsar's position.}
  \label{fig:smbh_spec}
\end{figure}

\begin{itemize}
\itemsep0em 
\item \textbf{Detection of a stochastic background of GWs from inspiraling SMBH binary systems}, thus proving that dynamical interactions in galactic centers can mitigate the ``final parsec problem'' of SMBH evolution, allowing SMBHs to merge in galaxy centers. \textit{The ngVLA has a goal resolution of 30 mas, equivalent to $\sim 300$ pc at $z\sim 1$, which is sufficient to distinguish dual AGN and convert observed fractions of dual/multiple AGN into SMBH merger rates. These rate estimates can inform future NANOGrav observing strategies.}
\item \textbf{Spectral characterization of the GW strain signal} which will lead to an inference of the dynamical processes influencing SMBH binaries at sub-parsec scales. The dominant process is expected to be scattering of stars on orbits that intersect the central galactic regions, or interaction with a viscous circumbinary disk.\textit{ If intercontinental VLBI capabilities are added to the ngVLA, it will achieve $\sim$pc-scale resolution of central galactic regions out to $z\sim0.1$. This will allow a large statistical sample of SMBH binary separations to be collected, thereby informing whether binaries stall at $\sim$pc separations, or are driven efficiently inward.}
\item \textbf{Constrain the scaling relationships between galactic bulge properties and central SMBHs} \citep{2000ApJ...539L...9F, 2000ApJ...539L..13G}, thereby illuminating the symbiotic evolution between the two, and shedding light on the growth of galaxies/SMBHs through cosmic time. \textit{The ngVLA will be able to resolve the influence radius of $10^9 M_\odot$ supermassive black holes out to $z\sim0.1$, drastically improving mass estimates that may lead to revisions of the $M_\mathrm{BH}-M_\mathrm{bulge}$ relationship.}
\item \textbf{Understanding how accretion of gas onto SMBHs in the galactic center is related to the inflow of gas at larger scales}. There is some question of whether accretion of material from a circumbinary disk is an important mechanism of SMBH binary hardening, given that the resident massive galaxies should in general be gas poor. But this is highly uncertain. \textit{ngVLA observations of gas reservoirs and inflow at larger scales, as well as gas dynamics in the innermost galaxy regions, will provide crucial constraints on gaseous accretion onto a central SMBH binary.}
\end{itemize}

\section{SMBH Dynamical Evolution In Post-merger Galaxies}

The dynamical evolution of SMBH binaries in post-merger galaxies is complex, involving a series of physical processes handing off to the next at ever smaller scales as the SMBHs become gravitationally bound, then tighten to milliparsec separations. We now describe the series of physical processes within post-merger galaxies that drive SMBHs to become gravitationally bound, and then to evolve as a binary into the nanohertz-emitting regime where PTAs are sensitive. At each stage, we comment on how ngVLA goals and observations interface with NANOGrav requirements.

\subsection{Dynamical friction} 

When massive galaxies merge together, their resident SMBHs sink to the center of the resulting galactic remnant through \textit{dynamical friction} \citep{merritt2005}. This is the consequence of many weak and long-range gravitational scattering events within the surrounding stellar, gas, and dark matter distributions, creating a drag that causes the SMBHs to decelerate and transfer energy to the ambient media \citep{chandrasekhar1943}. For systems with extreme mass ratios ($\lesssim 10^{-2}$) or very low total masses, dynamical friction may not be effective at forming a bound binary from the two SMBH within a Hubble time.  In this case the pair might become ``stalled'' at larger separations, with one of the two SMBH left to wander the galaxy at $\sim$ kpc separations \citep{db17}.  It is possible that a non-negligible fraction of galaxies may have such wandering SMBH, some of which may be observable as offset-AGN. 

\vspace{5mm}
\noindent$\quad\bullet\,\,$ If the ngVLA is expanded with intercontinental VLBI-like capabilities, the improvement in angular resolution over existing instruments will allow for direct imaging of systems that may contain off-set AGN to extend to much higher redshifts \citep{ngVLA8}, thereby potentially gathering a large enough statistical sample to test this stalling hypothesis.

\subsection{Stellar loss-cone scattering}
\label{sec:stars}

At parsec separations, dynamical friction hands off to individual $3$-body scattering events between stars in the galactic core and the SMBH binary \citep{bbr80}. Stars slingshot off the binary, which can extract orbital energy from the system \citep{mv92,q96}.
However, only stars in centrophilic orbits with very low angular momentum have trajectories which bring them deep enough into the galactic center to interact with the binary. The region of stellar-orbit phase space that is occupied by these types of stars is known as the ``loss cone" (LC) \citep{fr76}.  Stars which extract energy form the binary in a scattering event tend to be ejected from the core, depleting the LC. As with dynamical friction at larger scales, binaries can also stall here, at parsec scales, due to inefficiency of the LC-refilling ---a phenomenon referred to as the ``final parsec problem" \citep{mm02}.  Generally, binaries which don't reach sub-parsec separations will be unable to merge via GW emission within a Hubble time \citep{mop14}. 

Various mechanisms have been explored to see whether the LC can be efficiently refilled or populated to ensure continuous hardening of the binary down to milliparsec separations. In general, any form of bulge morphological triaxiality will ensure a continually-refilled LC that can mitigate the final-parsec problem \citep{vm13}. Isolated galaxies often exhibit triaxiality, and given that the SMBH binaries of interest are the result of galactic mergers, triaxiality and general asymmetries can be expected as a natural post-merger by-product. Also, post-merger galaxies often harbor large, dense molecular clouds that can be channeled into the galactic center, acting as a perturber for the stellar distribution that will refill the LC \citep{ys91}, or even directly hardening the binary \citep{gsc17}.

\vspace{5mm}
\noindent$\quad\bullet\,\,$ Current instruments cannot directly resolve molecular gas clouds within post-merger galaxies. However, the ngVLA will provide the sub-parsec resolution that is necessary to image these gas clouds \citep{ngVLA19}. This will provide a much-needed causal link explaining how the LC can remain filled in different kinds of post-merger galaxies.

\subsection{Viscous circumbinary disk interaction}
\label{sec:gas}

At centiparsec to milliparsec separations, viscous angular momentum exchange to a gaseous circumbinary disk may play an important role in hardening the binary \citep{bbr80, ka11}. This influence will depend on the details of the dissipative physics of the disk. The simplest model is that of a coplanar prograde $\alpha$-disk \citep{ss73}, where the binary torques the disk and carves out a cavity where no gas flows. The individual SMBHs may be surrounded by smaller ``mini-disks''. More complex disk structures are possible, such as the existence of several physically-distinct regions \citep{st86}. Additionally, high-density disks (equivalently: high-accretion rates) may provide rapid hardening, but are potentially unstable. Furthermore, hardening rates will change as the system passes through different types of migration based on the interaction between the secondary SMBH and the disk --- at larger separations the secondary SMBH is driven along with the gas flow, whereas when the mass enclosed by the binary orbit becomes comparable to the secondary SMBH the system is driven by binary dynamics. 

\vspace{5mm}
\noindent$\quad\bullet\,\,$ All of these complicated processes are happening on scales that are beyond the reach of current observational instruments. The ngVLA may have the capability to resolve the cold gas reservoirs feeding these disks and the subsequent AGN \citep{ngVLA40}. 

\subsection{Gravitational-wave inspiral}

The emission of gravitational radiation will always dominate binary orbital evolution at the smallest separation scales ($\lesssim$ milliparsec). This is when the binary has decoupled from the galactic environment and can be considered as an isolated physical system. In this case, the dissipation of orbital energy will depend only on the constituent SMBH masses, the orbital semi-major axis, and the binary's eccentricity.

\vspace{5mm}
\noindent$\quad\bullet\,\,$ This stage is below the angular resolution scale of the ngVLA, and will require PTA observations of nanohertz GWs to complete the picture of SMBH dynamical evolution.

\section{Anticipated Complementary Results}

NANOGrav's key goal is the detection and characterization of GWs at nanohertz frequencies, with the target being a population of inspiraling SMBH binaries. The stochastic gravitational wave background (GWB) signal produced by a population of inspiraling SMBH binaries can be described by its characteristic-strain spectrum, $h_c(f)$. The strain budget at different frequencies depends on various cosmological and astrophysical factors \citep{sesana2004}, which are summarized in Figure \ref{fig:smbh_spec} and in the following. In the simple case of a population of circular binaries whose orbital evolution is driven entirely by the emission of GWs, $h_c(f) \propto f^{-2/3}$.

Interaction of a binary with its surrounding galactic stellar distribution, as described in Section \ref{sec:stars}, will lead to low-frequency attenuation of the characteristic strain spectrum of GWs (i.e. a spectral turnover).  This can be separated into two distinct effects: $(1)$ the direct coupling leads to a faster reduction in the orbital separation than GW-driven inspiral, such that the amount of time spent by each binary at low frequencies is reduced; $(2)$ extraction of angular momentum by stellar slingshots can excite eccentricity, which leads to faster GW-driven inspiral, and (again) lower residence time at low frequencies. The excitation of binary eccentricity by interaction with stars will further attenuate the strain spectrum at low frequencies, leading to an even sharper turnover \citep[e.g.][]{tss17}. Coupling of a viscous circumbinary disk with a SMBH binary, as discussed in Section \ref{sec:gas}, will also lead to attenuation of the characteristic strain spectrum of GWs through both direct coupling, and excitation of eccentricity. 

\subsection{Detection of stochastic background}

NANOGrav expects to detect the nanohertz GWB within the next 2-3 years \citep{2016ApJ...819L...6T}. This initial detection will likely be of a long-timescale stochastic process, present in all pulsars and correlated between them with a signature unique to GWs. But it is unlikely that the shape of the spectrum will be very well constrained. Nevertheless, detection alone is sufficient to provide the first direct evidence that the final parsec problem of massive black-hole evolution is mitigated, and even with poorly constrained parameters, there will be many models of post-merger galactic environments that will be ruled out.

\subsection{Spectral characterization of background}

Over the next decade, NANOGrav's sensitivity will improve to the stage at which the slope of the GW strain spectrum will be well constrained. This will allow for robust inference of the dominant dynamical process for SMBH binary evolution. However, the entire population will consist of a mixture of galaxy types and dynamical environmental processes. Thus, it will only be through the combination of NANOGrav data with models built from the in-depth surveys of galactic cores, like those the ngVLA will conduct, that a complete understanding of the conditions under which the final parsec problem is mitigated will be gained.

\subsection{Constraining the scaling relationships between galaxies and central SMBHs}

NANOGrav limits on the stochastic GWB can already be expressed directly as limits on various model parameters, and the scaling relationship between galactic bulge properties and central SMBH masses has been shown to be one of the most important model parameters \citep{simonBS16,NG11}. In the coming years, NANOGrav will place tight constraints on the galaxy-SMBH scaling relationship, specifically for SMBHs in post-major-merger galaxies. As with constraints on the strain-spectrum slope, these constraints on the scaling relation can only be properly used to make a clear statement about galaxy growth when combined with electromagnetic observations of the associated galactic bulges. To this end, the ngVLA's proposed capabilities will allow resolution down to the influence radius of $10^9 M_\odot$ SMBHs out to $z\sim0.1$, which is more than sufficient to characterize properties of galactic bulges and to measure the SMBH mass. 

\subsection{Accretion onto SMBH}

NANOGrav's spectral characterization of the GWB signal will illuminate the important dynamical processes that harden SMBH binaries to milliparsec separations. This will allow a measurement of various astrophysical parameters, such as the mass density of loss-cone stars in galactic centers, binary eccentricities, and SMBH accretion rates. Inference of the latter will of course depend on the assumed disk dynamics, but will ultimately describe the relevance of gas dynamics are in this overall picture of SMBH dynamical evolution. It may be the case that the kinds of SMBHs that we are looking for reside in gas-poor massive galaxies, in which case stellar scattering will control the evolution at sub-parsec separations until GW-driven inspiral begins to dominate. We currently lack sufficient observational constraints on circumbinary disk dynamics to inform our intuition of its importance. The ngVLA will measure the accretion and build-up of H$1$ gas in galactic outskirts, track gas inflow to central regions, and (with expanded VLBI capabilities) be able to resolve down to the $\sim$pc scale out to $z\sim 0.1$. It will therefore give crucial insight into the importance of gas dynamics in SMBH evolution. 

\section{Summary \& Synergies}

Pulsar-timing arrays (such as NANOGrav) will detect the ensemble gravitational-wave signal from many inspiraling supermassive black-hole binaries throughout the Universe within the next $3-7$ years \citep{2016ApJ...819L...6T}. The statistical properties of this signal will reflect the dynamical history of these supermassive black-holes as they evolve to form a bound system and reach milliparsec orbital separations. NANOGrav will constrain the environments of supermassive black-hole binaries \citep{tss17} through detailed studies of the gravitational-wave strain spectrum in the nanohertz band. The ngVLA has a goal resolution of $30$ mas, equivalent to $\sim 300$ pc at $z\sim 1$. This resolution is sufficient to distinguish dual AGN, allowing estimates of SMBH merger rates to be formed that can act as prior constraints for PTA analysis. Furthermore, if intercontinental VLBI capabilities are added to the ngVLA it will achieve $\sim$pc-scale resolution out to $z\sim0.1$, allowing a large measured sample of SMBH binary separations to inform whether binaries stall at $\sim$pc separations or are driven efficiently to the sub-pc regime by dynamical interactions. Intercontinental VLBA will also allow the influence radius of supermassive black-holes to be resolved, which will drastically improve mass estimation and thus the fidelity of derived $M_\mathrm{BH}-M_\mathrm{bulge}$ relationships. Finally, ngVLA observations of galactic gas reservoirs, large-scale inflows, and gas dynamics in central galactic regions will provide crucial insight into the physics of SMBH binary accretion processes. 

There are several current and future facilities that will complement the ngVLA in the aforementioned areas. These include the \textit{Atacama Large Millimeter/submillimeter Array} (ALMA)\footnote{http://www.eso.org/public/usa/teles-instr/alma}, whose angular resolution at $230$ GHz with baseline length of $15$ km is $15$ mas, which is comparable to the ngVLA's goal resolution of $30$ mas at $40$ GHz with a baseline length of $180$ km \citep{ngVLA8}. Likewise, the ngVLA should have similar angular resolution to adaptive-optics corrected images from the \textit{Giant Magellan Telescope} (GMT)\footnote{\href{https://www.gmto.org}{https://www.gmto.org}}, \textit{Thirty Meter Telescope} (TMT)\footnote{\href{https://www.tmt.org}{https://www.tmt.org}}, and the \textit{Extremely Large Telescope} (ELT)\footnote{https://www.eso.org/public/usa/teles-instr/elt}. This range of complementarity in angular resolution will allow for validation and comparison studies of sub-kiloparsec galaxy scales out to $z\sim 1$, which is sufficient to distinguish dual AGN for the purpose of estimating SMBH binary merger rates. On the other hand, the SKA\footnote{\href{https://www.skatelescope.org}{https://www.skatelescope.org}} is expected to operate concurrently with the ngVLA, but its point-source sensitivity will be only be half that of the ngVLA in their frequency-overlap band \citep{ngVLA8}. Similarly, JWST\footnote{\href{https://www.jwst.nasa.gov}{https://www.jwst.nasa.gov}} will operate at similar wavelengths to GMT/TMT/ELT, but have lower angular resolution than those instruments and the ngVLA.

NANOGrav gravitational-wave analysis, in concert with observations by the ngVLA and complementary facilities, will paint a multi-messenger portrait of galaxy evolution and the dynamics of the most massive black-holes in the Universe.

\acknowledgements The NANOGrav project receives support from NSF Physics Frontier Center award number 1430284. SRT acknowledges support from the NANOGrav Physics Frontier Center. JS acknowledge support from the JPL RTD program. Some of this research was undertaken at the Jet Propulsion Laboratory, California Institute of Technology, under a contract with the National Aeronautics and Space Administration.

\bibliography{references}

\begin{thebibliography}{}
\expandafter\ifx\csname natexlab\endcsname\relax\def\natexlab#1{#1}\fi
\expandafter\ifx\csname url\endcsname\relax
  \def\url#1{\texttt{#1}}\fi
\expandafter\ifx\csname urlprefix\endcsname\relax\def\urlprefix{URL }\fi
\providecommand{\eprint}[2][]{\url{#2}}

\bibitem[{{Arzoumanian} et~al.(2018){Arzoumanian}, {Baker}, {Brazier},
  {Burke-Spolaor}, {Chamberlin}, {Chatterjee}, {Christy}, {Cordes}, {Cornish},
  {Crawford}, {Thankful Cromartie}, {Crowter}, {DeCesar}, {Demorest}, {Dolch},
  {Ellis}, {Ferdman}, {Ferrara}, {Folkner}, {Fonseca}, {Garver-Daniels},
  {Gentile}, {Haas}, {Hazboun}, {Huerta}, {Islo}, {Jones}, {Jones}, {Kaplan},
  {Kaspi}, {Lam}, {Lazio}, {Levin}, {Lommen}, {Lorimer}, {Luo}, {Lynch},
  {Madison}, {McLaughlin}, {McWilliams}, {Mingarelli}, {Ng}, {Nice}, {Park},
  {Pennucci}, {Pol}, {Ransom}, {Ray}, {Rasskazov}, {Siemens}, {Simon},
  {Spiewak}, {Stairs}, {Stinebring}, {Stovall}, {Swiggum}, {Taylor},
  {Vallisneri}, {van Haasteren}, {Vigeland}, {Zhu}, \& {The NANOGrav
  Collaboration}}]{NG11}
{Arzoumanian}, Z., {Baker}, P.~T., {Brazier}, A., {Burke-Spolaor}, S.,
  {Chamberlin}, S.~J., {Chatterjee}, S., {Christy}, B., {Cordes}, J.~M.,
  {Cornish}, N.~J., {Crawford}, F., {Thankful Cromartie}, H., {Crowter}, K.,
  {DeCesar}, M., {Demorest}, P.~B., {Dolch}, T., {Ellis}, J.~A., {Ferdman},
  R.~D., {Ferrara}, E., {Folkner}, W.~M., {Fonseca}, E., {Garver-Daniels}, N.,
  {Gentile}, P.~A., {Haas}, R., {Hazboun}, J.~S., {Huerta}, E.~A., {Islo}, K.,
  {Jones}, G., {Jones}, M.~L., {Kaplan}, D.~L., {Kaspi}, V.~M., {Lam}, M.~T.,
  {Lazio}, T.~J.~W., {Levin}, L., {Lommen}, A.~N., {Lorimer}, D.~R., {Luo}, J.,
  {Lynch}, R.~S., {Madison}, D.~R., {McLaughlin}, M.~A., {McWilliams}, S.~T.,
  {Mingarelli}, C.~M.~F., {Ng}, C., {Nice}, D.~J., {Park}, R.~S., {Pennucci},
  T.~T., {Pol}, N.~S., {Ransom}, S.~M., {Ray}, P.~S., {Rasskazov}, A.,
  {Siemens}, X., {Simon}, J., {Spiewak}, R., {Stairs}, I.~H., {Stinebring},
  D.~R., {Stovall}, K., {Swiggum}, J., {Taylor}, S.~R., {Vallisneri}, M., {van
  Haasteren}, R., {Vigeland}, S., {Zhu}, W.~W., \& {The NANOGrav Collaboration}
  2018, \apj, 859, 47. \eprint{1801.02617}

\bibitem[{{Begelman} et~al.(1980){Begelman}, {Blandford}, \& {Rees}}]{bbr80}
{Begelman}, M.~C., {Blandford}, R.~D., \& {Rees}, M.~J. 1980, \nat, 287, 307

\bibitem[{{Blumenthal} et~al.(1984){Blumenthal}, {Faber}, {Primack}, \&
  {Rees}}]{1984Natur.311..517B}
{Blumenthal}, G.~R., {Faber}, S.~M., {Primack}, J.~R., \& {Rees}, M.~J. 1984,
  \nat, 311, 517

\bibitem[{{Bolatto} et~al.(2017){Bolatto}, {Chatterjee}, {Casey}, {Chomiuk},
  {de Pater}, {Dickinson}, {Di Francesco}, {Hallinan}, {Isella}, {Kohno},
  {Kulkarni}, {Lang}, {Lazio}, {Leroy}, {Loinard}, {Maccarone}, {Matthews},
  {Osten}, {Reid}, {Riechers}, {Sakai}, {Walter}, \& {Wilner}}]{ngVLA19}
{Bolatto}, A.~D., {Chatterjee}, S., {Casey}, C.~M., {Chomiuk}, L., {de Pater},
  I., {Dickinson}, M., {Di Francesco}, J., {Hallinan}, G., {Isella}, A.,
  {Kohno}, K., {Kulkarni}, S.~R., {Lang}, C., {Lazio}, T.~J.~W., {Leroy},
  A.~K., {Loinard}, L., {Maccarone}, T.~J., {Matthews}, B.~C., {Osten}, R.~A.,
  {Reid}, M.~J., {Riechers}, D., {Sakai}, N., {Walter}, F., \& {Wilner}, D.
  2017, ArXiv e-prints. \eprint{1711.09960}

\bibitem[{{Casey} et~al.(2015){Casey}, {Hodge}, {Lacy}, {Hales}, {Barger},
  {Narayanan}, {Carilli}, {Alatalo}, {da Cunha}, {Emonts}, {Ivison}, {Kimball},
  {Kohno}, {Murphy}, {Riechers}, {Sargent}, \& {Walter}}]{ngVLA8}
{Casey}, C.~M., {Hodge}, J.~A., {Lacy}, M., {Hales}, C.~A., {Barger}, A.,
  {Narayanan}, D., {Carilli}, C., {Alatalo}, K., {da Cunha}, E., {Emonts}, B.,
  {Ivison}, R., {Kimball}, A., {Kohno}, K., {Murphy}, E., {Riechers}, D.,
  {Sargent}, M., \& {Walter}, F. 2015, NRAO Next Generation Very Large Array
  Memos Series. \eprint{1510.06411},
  \urlprefix\url{http://library.nrao.edu/ngvla.shtml}

\bibitem[{{Chandrasekhar}(1943)}]{chandrasekhar1943}
{Chandrasekhar}, S. 1943, \apj, 97, 255

\bibitem[{{Dvorkin} \& {Barausse}(2017)}]{db17}
{Dvorkin}, I., \& {Barausse}, E. 2017, \mnras, 470, 4547. \eprint{1702.06964}

\bibitem[{{Ferrarese} \& {Merritt}(2000)}]{2000ApJ...539L...9F}
{Ferrarese}, L., \& {Merritt}, D. 2000, \apjl, 539, L9.
  \eprint{astro-ph/0006053}

\bibitem[{Frank \& Rees(1976)}]{fr76}
Frank, J., \& Rees, M.~J. 1976, Monthly Notices of the Royal Astronomical
  Society, 176, 633

\bibitem[{{Gebhardt} et~al.(2000){Gebhardt}, {Bender}, {Bower}, {Dressler},
  {Faber}, {Filippenko}, {Green}, {Grillmair}, {Ho}, {Kormendy}, {Lauer},
  {Magorrian}, {Pinkney}, {Richstone}, \& {Tremaine}}]{2000ApJ...539L..13G}
{Gebhardt}, K., {Bender}, R., {Bower}, G., {Dressler}, A., {Faber}, S.~M.,
  {Filippenko}, A.~V., {Green}, R., {Grillmair}, C., {Ho}, L.~C., {Kormendy},
  J., {Lauer}, T.~R., {Magorrian}, J., {Pinkney}, J., {Richstone}, D., \&
  {Tremaine}, S. 2000, \apjl, 539, L13. \eprint{astro-ph/0006289}

\bibitem[{{Goicovic} et~al.(2017){Goicovic}, {Sesana}, {Cuadra}, \&
  {Stasyszyn}}]{gsc17}
{Goicovic}, F.~G., {Sesana}, A., {Cuadra}, J., \& {Stasyszyn}, F. 2017, \mnras,
  472, 514. \eprint{1602.01966}

\bibitem[{{Kocsis} \& {Sesana}(2011)}]{ka11}
{Kocsis}, B., \& {Sesana}, A. 2011, \mnras, 411, 1467. \eprint{1002.0584}

\bibitem[{{Kormendy} \& {Richstone}(1995)}]{kormendy1995}
{Kormendy}, J., \& {Richstone}, D. 1995, \araa, 33, 581

\bibitem[{{Magorrian} et~al.(1998){Magorrian}, {Tremaine}, {Richstone},
  {Bender}, {Bower}, {Dressler}, {Faber}, {Gebhardt}, {Green}, {Grillmair},
  {Kormendy}, \& {Lauer}}]{Magorrian1998}
{Magorrian}, J., {Tremaine}, S., {Richstone}, D., {Bender}, R., {Bower}, G.,
  {Dressler}, A., {Faber}, S.~M., {Gebhardt}, K., {Green}, R., {Grillmair}, C.,
  {Kormendy}, J., \& {Lauer}, T. 1998, \aj, 115, 2285.
  \eprint{astro-ph/9708072}

\bibitem[{{McWilliams} et~al.(2014){McWilliams}, {Ostriker}, \&
  {Pretorius}}]{mop14}
{McWilliams}, S.~T., {Ostriker}, J.~P., \& {Pretorius}, F. 2014, \apj, 789,
  156. \eprint{1211.5377}

\bibitem[{{Merritt} \& {Milosavljevi{\'c}}(2005)}]{merritt2005}
{Merritt}, D., \& {Milosavljevi{\'c}}, M. 2005, Living Reviews in Relativity,
  8. \eprint{astro-ph/0410364}

\bibitem[{{Mikkola} \& {Valtonen}(1992)}]{mv92}
{Mikkola}, S., \& {Valtonen}, M.~J. 1992, \mnras, 259, 115

\bibitem[{{Milosavljevi{\'c}} \& {Merritt}(2003)}]{mm02}
{Milosavljevi{\'c}}, M., \& {Merritt}, D. 2003, in The Astrophysics of
  Gravitational Wave Sources, edited by J.~M. {Centrella}, vol. 686 of American
  Institute of Physics Conference Series, 201. \eprint{astro-ph/0212270}

\bibitem[{{Nyland} et~al.(2018){Nyland}, {Harwood}, {Mukherjee}, {Jagannathan},
  {Rujopakarn}, {Emonts}, {Alatalo}, {Bicknell}, {Davis}, {Greene}, {Kimball},
  {Lacy}, {Lonsdale}, {Lonsdale}, {Maksym}, {Molnar}, {Morabito}, {Murphy},
  {Patil}, {Prandoni}, {Sargent}, \& {Vlahakis}}]{ngVLA40}
{Nyland}, K., {Harwood}, J.~J., {Mukherjee}, D., {Jagannathan}, P.,
  {Rujopakarn}, W., {Emonts}, B., {Alatalo}, K., {Bicknell}, G., {Davis},
  T.~A., {Greene}, J., {Kimball}, A., {Lacy}, M., {Lonsdale}, C., {Lonsdale},
  C., {Maksym}, W.~P., {Molnar}, D., {Morabito}, L., {Murphy}, E., {Patil}, P.,
  {Prandoni}, I., {Sargent}, M., \& {Vlahakis}, C. 2018, ArXiv e-prints.
  \eprint{1803.02357}

\bibitem[{{Quinlan}(1996)}]{q96}
{Quinlan}, G.~D. 1996, New Astronomy, 1, 35. \eprint{astro-ph/9601092}

\bibitem[{{Sesana} et~al.(2004){Sesana}, {Haardt}, {Madau}, \&
  {Volonteri}}]{sesana2004}
{Sesana}, A., {Haardt}, F., {Madau}, P., \& {Volonteri}, M. 2004, \apj, 611,
  623. \eprint{astro-ph/0401543}

\bibitem[{{Shakura} \& {Sunyaev}(1973)}]{ss73}
{Shakura}, N.~I., \& {Sunyaev}, R.~A. 1973, \aap, 24, 337

\bibitem[{{Shapiro} \& {Teukolsky}(1986)}]{st86}
{Shapiro}, S.~L., \& {Teukolsky}, S.~A. 1986, {Black Holes, White Dwarfs and
  Neutron Stars: The Physics of Compact Objects} (John Wiley \& Sons)

\bibitem[{{Simon} \& {Burke-Spolaor}(2016)}]{simonBS16}
{Simon}, J., \& {Burke-Spolaor}, S. 2016, \apj, 826, 11. \eprint{1603.06577}

\bibitem[{{Taylor} et~al.(2017){Taylor}, {Simon}, \& {Sampson}}]{tss17}
{Taylor}, S.~R., {Simon}, J., \& {Sampson}, L. 2017, Physical Review Letters,
  118, 181102. \eprint{1612.02817}

\bibitem[{{Taylor} et~al.(2016){Taylor}, {Vallisneri}, {Ellis}, {Mingarelli},
  {Lazio}, \& {van Haasteren}}]{2016ApJ...819L...6T}
{Taylor}, S.~R., {Vallisneri}, M., {Ellis}, J.~A., {Mingarelli}, C.~M.~F.,
  {Lazio}, T.~J.~W., \& {van Haasteren}, R. 2016, \apjl, 819, L6.
  \eprint{1511.05564}

\bibitem[{{Vasiliev} \& {Merritt}(2013)}]{vm13}
{Vasiliev}, E., \& {Merritt}, D. 2013, \apj, 774, 87. \eprint{1301.3150}

\bibitem[{{Young} \& {Scoville}(1991)}]{ys91}
{Young}, J.~S., \& {Scoville}, N.~Z. 1991, \araa, 29, 581

\end{thebibliography}

\end{document}